\documentclass[a4paper]{article}
\usepackage[margin=1.5in]{geometry}

\usepackage{authblk}
\usepackage[utf8]{inputenc}
\usepackage{amsmath,amssymb,color,cite,hyperref,graphicx,caption,subcaption}
\usepackage{amsthm}
\usepackage{booktabs} 
\usepackage[ruled]{algorithm2e} 
\usepackage{enumitem}

\newcommand{\R}{{\mathbb{R}}}
\newcommand{\N}{{\mathbb{N}}}

\newcommand{\C}{{\mathbb{C}}}

\theoremstyle{definition}
\newtheorem{definition}{Definition}[section] 
\theoremstyle{remark}
\newtheorem*{remark}{Remark}

\begin{document}

\title{Introducing Data Primitives:  \\
	Data Formats for the SKED Framework}

\author[1]{Elizabeth D. Trippe}

\author[2]{Jacob B. Aguilar}

\author[1]{Yi H. Yan}

\author[3]{Mustafa V. Nural}

\author[4]{Jessica A. Brady}

\author[1,2,3]{Juan B. Gutierrez \footnote{Corresponding author: jgutierr@uga.edu}}

  \affil[1]{Institute of Bioinformatics, University of Georgia}
  \affil[2]{Department of Mathematics, University of Georgia}
  \affil[3]{Department of Computer Science, University of Georgia}
  \affil[4]{School of Engineering, University of Georgia}

\maketitle

\begin{abstract} 
\textbf{\large Background}\\

\noindent The past few years have seen a tremendous increase in the size and complexity of datasets.  Scientific and clinical studies must to incorporate datasets that cross multiple spatial and temporal scales to describe a particular phenomenon. The storage and accessibility of these heterogeneous datasets in a way that is useful to researchers and yet extensible to new data types is a major challenge. \\
 
\textbf{\large Methods}\\

\noindent In order to overcome these obstacles, we propose the use of data primitives as a common currency between analytical methods.  The four data primitives we have identified are time series, text, annotated graph and triangulated mesh, with associated metadata.  Using only data primitives to store data and as algorithm input, output, and intermediate results, promotes interoperability, scalability, and reproducibility in scientific studies. \\
 
\textbf{\large Results}\\

\noindent Data primitives were used in a multi-omic, multi-scale systems biology study of malaria infection in non-human primates to perform many types of integrative analysis quickly and efficiently.  \\

\textbf{\large Conclusions}\\

\noindent Using data primitives as a common currency for both data storage and for cross talk between analytical methods enables the analysis of complex multi-omic, multi-scale datasets in a reproducible modular fashion.\\

\end{abstract}


\section{Background}\label{sec:Backgrd}

 		Recent advances in data acquisition have resulted in a rapid increase in data types and models for the study of systems biology.  Models in biology describe our current understanding of a system and provide the theoretical basis for further experiments and investigations \cite{drager_improving_2014}.	Because biological models have become increasingly complex over the years with even models to describe an entire cell, it is impractical to rely on traditional methods (ie. published papers, books) to contain all necessary information  to recreate an experiment and design future experiments.  A quantitative model using ordinary differential equations (ODEs) may include hundreds of equations and still not adequately describe a biological phenomenon. For example, muscular dystrophy (MD) models combine the presence of physiological symptoms and cellular quantities to describe the progression of the disease \cite{2016arXiv161003521C}. Such models are essential in predicting disease outcomes \cite{2016arXiv161003521C}. 

 		Previous standardization efforts for model sharing in systems biology are limited in scope and can require a great deal of human interpretation. Standardization efforts have been focused on one level of biological organization and even systems biology standardization efforts are cell -centric with few resources to connect such models to clinical or physiological data and even simulation experiments have their own standardization format \cite{dada_sbrml-_2011}.  To use current models for an integrated analysis, multiple different standardization schemes are required and there are standards for storing and exchanging all the necessary information \cite{bergmann2014combine}.  \\
 		
 		While focus on individual levels of biological organization has lead to many useful standards, it is not an efficient approach to enable large scale multi-omic data integration. This analysis requires simultaneous integration of multiple variables from different standard data formats.  Even with the use of automated tools and established workflows, this process can be cumbersome and laborious.  New data formats are generated by the creation of new technologies (ex. ChIP-seq data) and the number of standard data formats continues to increase.\\
 		
 		Next, the extension of systems biology tools and results to systems biological engineering or synthetic biology has been hampered by a lack of comprehensive predictive models.  Even in the design of a minimal bacterial genome, 17\% of necessary genes could not be assigned known functions \cite{Hutchisonaad6253}. Clearly more accurate descriptions are needed to enable efficient manipulations of biological systems and to predict outcomes of economic interest. \\

\begin{figure}[ht!]
			\centering
			\begin{subfigure}[t]{.8\textwidth}
				\centering
				\includegraphics[width=0.8\linewidth]{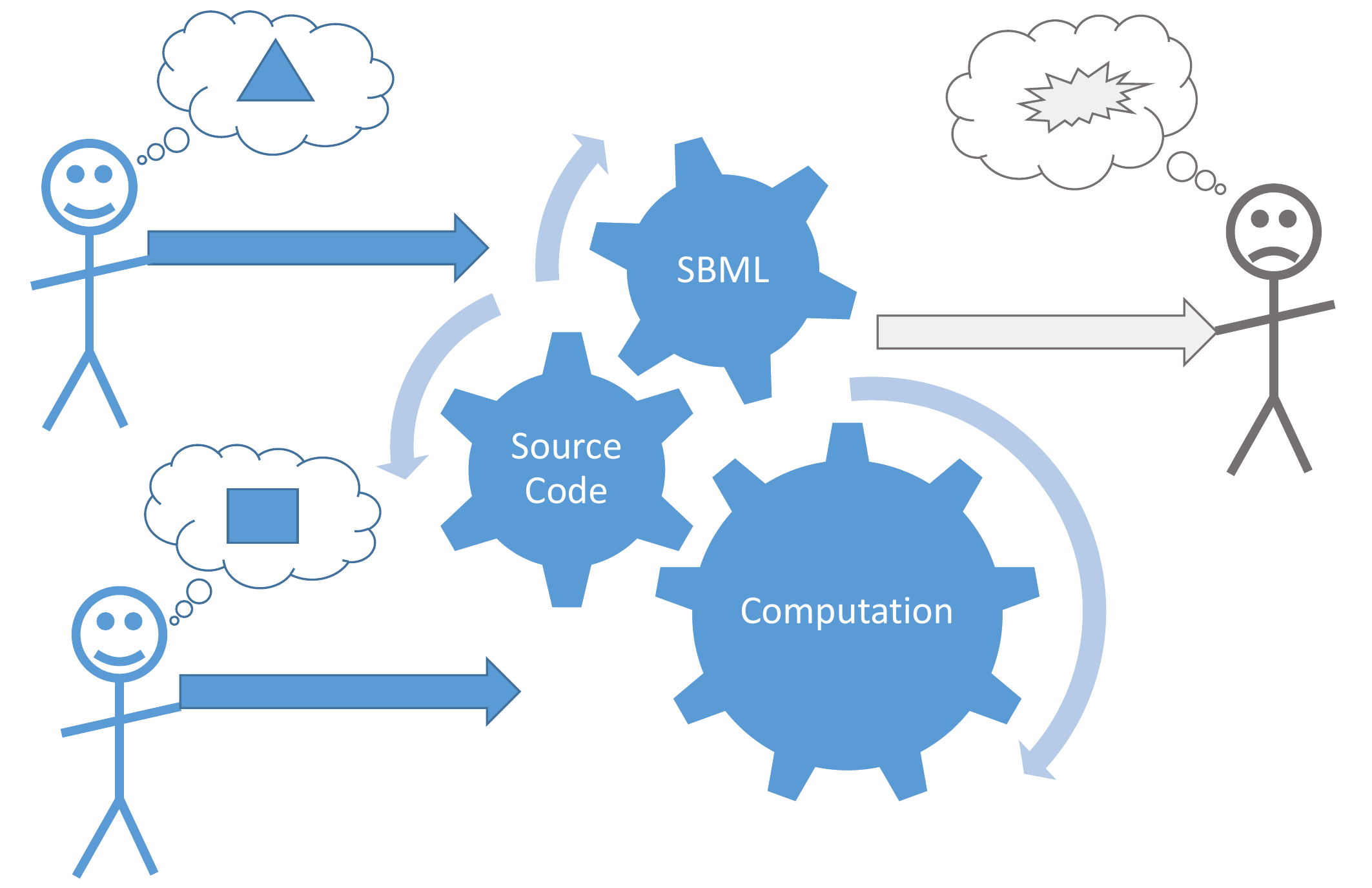}
				\caption{Sharing source code and model definition limits the use of a model to identical situations. A scientist with a fundamental different problem (gray icon) cannot reuse a model developed in a different context.}
				\label{fig:ComparisonA}
			\end{subfigure}%
			\vspace{0.2in}
			\begin{subfigure}[t]{.8\textwidth}
				\centering
				\includegraphics[width=0.8\linewidth]{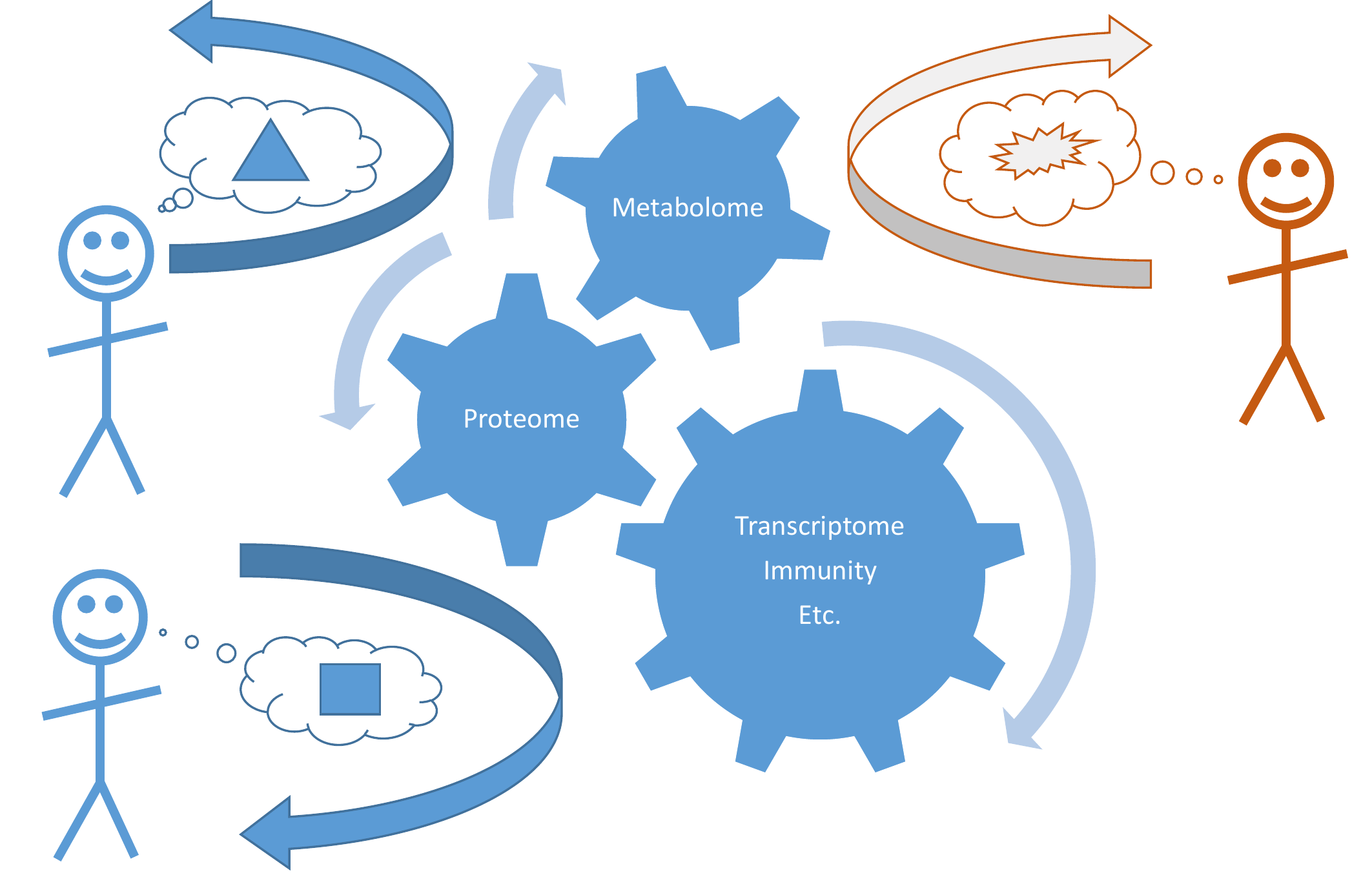}
				\caption{Algorithms and their implementation can be reused in different contexts if they use the same data primitives; the trust on a model might increase if it has been validated with a gold standard.}
				\label{fig:ComparisonB}
			\end{subfigure}
			\caption{Comparison of code sharing vs. data primitives}
			\label{fig:Comparison}
		\end{figure}

		The interoperability of models can occur in a variety of ways. Figure \ref{fig:Comparison} depicts two scenarios of model sharing: (a) source code and/or an SBML representation is used to capture the definition of a model; however, these models could not be used in contexts different from the specific problem for which they were created, and (b) quantitative scientists share algorithms that have been validated with a data set of known properties; however, sharing algorithms, their implementation, and their results is a non-trivial task unless the inputs and outputs of an analytic pipeline are standardized.  
		
		An elegant scalable solution to harmonize multi-omic data does not currently exist; while there are already resources in place that can and do make data more accessible, few, if any make it usable to address broad questions without substantial effort. The US National Institutes of Health supports the National Center for Biotechnology Information, which has a collection of resources (GeneBank, OMIM, MMDB, UniGene, Entrez, etc.). NIH also supports four Bioinformatics Resource Centers resulting in 6 community resources: (1) EupathDB focusing in eukaryotic parasites \cite{aurrecoechea2010eupathdb,aurrecoechea2012eupathdb}, (2) FungiDB focusing on fungal pathogens \cite{stajich2011fungidb}, (3) VectorBase focusing on invertebrate vectors of infectious disease \cite{lawson2007vectorbase,lawson2009vectorbase,megy2012vectorbase}, (4) ViPR focusing on viruses \cite{pickett2012virus,pickett2012vipr}, (5) IRD focusing exclusively on influenza \cite{squires2012influenza}, and (6) PATRIC focusing on bacterial pathogens \cite{snyder2007patric,wattam2013patric}. Even though all these resources could in principle accommodate multi-omic and clinical data, each one is restricted by their own data access architecture and focus. Therefore, a quantitative model implemented in one database cannot be easily made available to other databases, even if the data consumed by the model is fundamentally the same.\\
		
		Next, reproducibility of computational research has been identified as one challenge for systems biology \cite{peng2011reproducible,sandve2013ten}. When reproducing computational results, "forensic bioinformatics, where a scientist must check the input and output data to determine the methods that have been used, must often be used when documentation and directions did not provide enough information \cite{baggerly2009}. One case study describes  a novice user needing ~280 hours to reproduce a method \cite{garijo_quantifying_2013}.  With the fast pace of research and the need to make the most of valuable high-throughput experimental results, computational findings need to be reliable and easy to use \cite{prokop_reproducibility_2013}.\\

\begin{figure}[ht]
 		\centering
 		\includegraphics[width=0.8\linewidth]{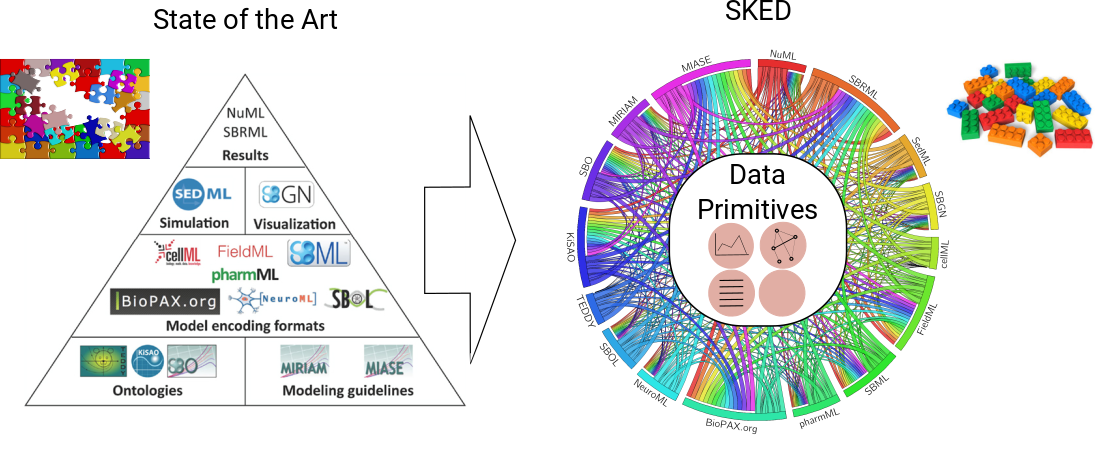}
 		\caption{ Data Primitives simplify how standards talk with each other (pyramid modified from Drager and Palsson, 2014 \cite{drager2014improving}, Figure reproduced from "A Vision for Health Informatics" submitted to KDD 2017 BigDAS Workshop)  }%
 		\label{fig:StdstoDP} 
 	\end{figure}

 In order to overcome these obstacles, we propose the describe the Scientific Knowledge Extraction from Data (SKED) framework and describe here the use of data primitives as a common currency between analytical methods, thus promoting interoperability, scalability, and reproducibility.  Figure \ref{fig:StdstoDP} describes this paradigmatic shift from many data standards to four data primitives which are inter-converted during analysis.  Using data primitives enables a single analytical method to be seamlessly utilized for different contexts.  Even though it has been stated that there is no "one-size-fits-all" standard for biological research\cite{drager_improving_2014} and "top-down standards will not serve systems biology"\cite{quackenbush2006top}, we believe that our efforts are fundamentally different and that many previous efforts may be combined through the use of data primitives.
 
\section{Methods}\label{sec:Methods}
Listed below is an overview of the notation and mathematical framework utilized throughout this work. Firstly, let $\N:=\{0,1,2,...\}$ be the set of natural numbers, so that $\N_+:=\N-\{0\}$ denotes the maximal strictly positive subset of the natural numbers. Additionally, for $n \in \N_+$ let $\R^n$ be the finite dimensional vector space comprised of $n$-tuples of real numbers. Let $\C:=\{\bar{z} \hspace{3pt}:\hspace{3pt} \bar{z}=a+bi \hspace{3pt}\text{where}\hspace{3pt} (a,b) \in \R^2\hspace{3pt}\text{and}\hspace{3pt} i=\sqrt{-1} \}$ be the field of complex numbers, so that for $m \in \N_+$ the symbol $\C^m$ denotes the space of complex valued $m$-tuples. Furthermore, for $m,n \in \N_+$ denote the set of $m \times n$ real valued matrices by $\mathbb{R}^{m \times n}$. Finally, interval notation is to be interpreted with respect to the underlying ordering (if any) imposed on the elements in the interval. 

We define \textit{data primitives} as a limited set of data structures that could serve as uniform building blocks to access information across different data sources. The concept of data primitives is analogous to using LEGO\textregistered\space blocks to build analysis pipelines, and allow re-purposing and redesign of quantitative methods with little overhead. 

{
}
\theoremstyle{definition}
\begin{definition}{\textbf{Time Series.}}
Let $i, m, n \in \N$, a \textit{time series} consisting of $n$ time points and $m$ variables is a totally ordered set $X := \{ (t_i, x_i)\}$, such that $x_i \in \C^m$ for $i \in [0,n]$.
\end{definition}
\theoremstyle{definition}
\begin{definition}{\textbf{Graph.}}
A triple $G=(V,E,W)$ is called a \textit{graph} on $V$, where:
\begin{enumerate}[label=\Roman*]
	\item It is assumed that $V$ is totally ordered and countable such that $|V|=m$, for $m \in \N_+$. The restriction that $m \in \N_+$ ensures that the set $V$ is nonempty, i.e. $V \neq \varnothing$. 
With respect to a particular ordering, the set $V$ can be uniquely represented as $V = \{ v_1, v_2, ...,v_m \}$. The set $V$ is called the \textit{vertex set} and each $v_i\in V$ is known as a \textit{vertex}. 	
	\item For $n \in \N$ and $i,j,k\in \N_+$ such that $k \in [1,n]$, the \textit{edge set} $E:=\{e_k:=(v_i,v_j) \in V \times V \hspace{3pt}:\hspace{3pt} v_i \neq v_j\}$ is comprised of $n$ unordered tuples called \textit{edges}. It is worth mentioning that $i,j \in [1,m]$ where $i \prec j \preceq m$ and/or $j \prec i \preceq m $. Also, the case $n=0$	corresponds to $E=\varnothing$, i.e. no edges are present in the graph under consideration.
	\item Let the symbol $\varphi$ denote a correspondence which assigns each element in $E$ to an unordered $q$-tuple $\omega_k \in \R^q$, where $q \in \N_+$. Denote the set $W \subset \R^q$ to be the image of $E$ under the mapping $\varphi$. In this case, each $\omega_k \in W$ represents a set of numeric values associated with each edge, presenting e.g. distance, capacity, weight, etc. The assumption is made that $\varphi$ is a surjection, i.e. for all $\omega_k \in W$, there exists an edge $e_k \in E$ such that $\omega_k=\varphi(e_k)$, where $k \in [1,n]$. In set notation it follows that $\varphi:E \twoheadrightarrow W$. 
\end{enumerate}
\end{definition}
 The graph $G$ has an associated \textit{adjacency matrix} $A \in \mathbb{R}^{m^2}$ with elements given by
$$
  a_{i,j}=
  \begin{cases}
  1, \mbox{ if } (v_i,v_j) \in E\\
  0, \mbox{ otherwise.}
  \end{cases}
$$
\theoremstyle{definition}
\begin{definition}{\textbf{Triangulated Mesh.}}
A triple $\mathcal{T}=(V,E,T)$ is called a \textit{triangulated mesh}, provided that the following three conditions are satisfied.
\begin{enumerate}[label=\Roman*]
	\item Let $l \in [1,m]$, then for all vertices $v_l \in V$, there is an edge $(v_i,v_j)\in E$ such that $v_l=v_i$ $\vee$ $v_l=v_j$. 
	\item For $p \in \N_+$ and $s \in [1,p]$, define the set $T:=\{\tau_s:=(v_i,v_j,v_k) \in V \times V \times V \hspace{3pt}:\hspace{3pt} v_i \neq v_j \neq v_k\}$. The set $T$ is composed of $p$ unordered triples called \textit{triangles}. Making use of a slight abuse of notation, it is required that for all $(v_i,v_j)\in E$, there exists a triangle $\tau_s=(v_i,v_j,v_k) \in T$ such that $(v_i,v_j)\in \tau_s$. 
	\item Provided two triangles intersect, i.e. $\tau_r \cap \tau_s \neq \varnothing$, then the vertex or edge responsible for the nonempty intersection is contained in $\mathcal{T}$.
\end{enumerate}
\end{definition}

\theoremstyle{definition}
\begin{definition}{\textbf{Text.}}
A string of characters is  stored as text and is used in qualitative descriptions.  It may or may not be a part of Metadata, described below. 
\end{definition}
\theoremstyle{definition}
\begin{definition}{\textbf{Metadata.}}
Let the sets of \textit{total data}, \textit{meta data} and \textit{data} obtained from a given experiment be labeled by $S_T$, $S_M$ and $S_D$, respectively. The set $S_T$
admits a unique mutually disjoint decomposition with respect to the analysis conducted, i.e. $S_T=S_M \cup S_D$ where $S_M \cap S_D= \varnothing$. Elements of $S_M$ can be thought of as data that provides information about experimentally obtained data, e.g. the instruments used, the instrument operators, dates, etc. For $n \in \N_+$ let $\Phi:=\{\phi_1,...,\phi_n \}$ be a family of mappings such that for each $s_{m_i} \in S_M$
and $S_{D_i} \subset S_D$, we have that $\phi_i(S_{D_i})=s_{m_i}$. In this case, all of the sets under consideration are countable and finite, as a result 
\begin{equation*}
	S_M=\bigcup_{i=1}^n \phi_i(S_{D_i}).
\end{equation*}
	\label{MetaDef}
\end{definition}
\begin{remark}
A triangulated mesh is a particular case of more general structures called simplices. A $k$-simplex is a $k$-dimensional geometric object with flat sides which is the convex hull of its $k + 1$ vertices. The mesh stores the vertex, edge and face information of a given surface or data set and is a piecewise planar surface, i.e. it is planar almost everywhere, except at the edges where the triangles join. In the case where all of the faces are triangles, the mesh is called triangulated. Therefore, a triangulated mesh can be regarded as a collection of triangles in three dimensional space that are connected in a particular way (to form a manifold on the given surface, i.e. each edge is shared by no more than two faces). It is well known that any surface can be estimated by a series of triangles. Each triangle can store additional data at the faces, e.g. colors, with sharp creases stored on edges and continuously varying quantities stored at each vertex. Due to their relatively simple geometric structure, all triangles can be represented as triples. An advantage of using such a mesh lies in the ability to efficiently answer data queries (information requests from a given database), e.g. finding the vertices or edges of a particular face or finding all triangles around a vertex.
\end{remark}

\section{Results and Discussion}\label{sec:ResultsAndDisc}

\begin{figure}[ht]
 		\centering
 		\includegraphics[width=\linewidth]{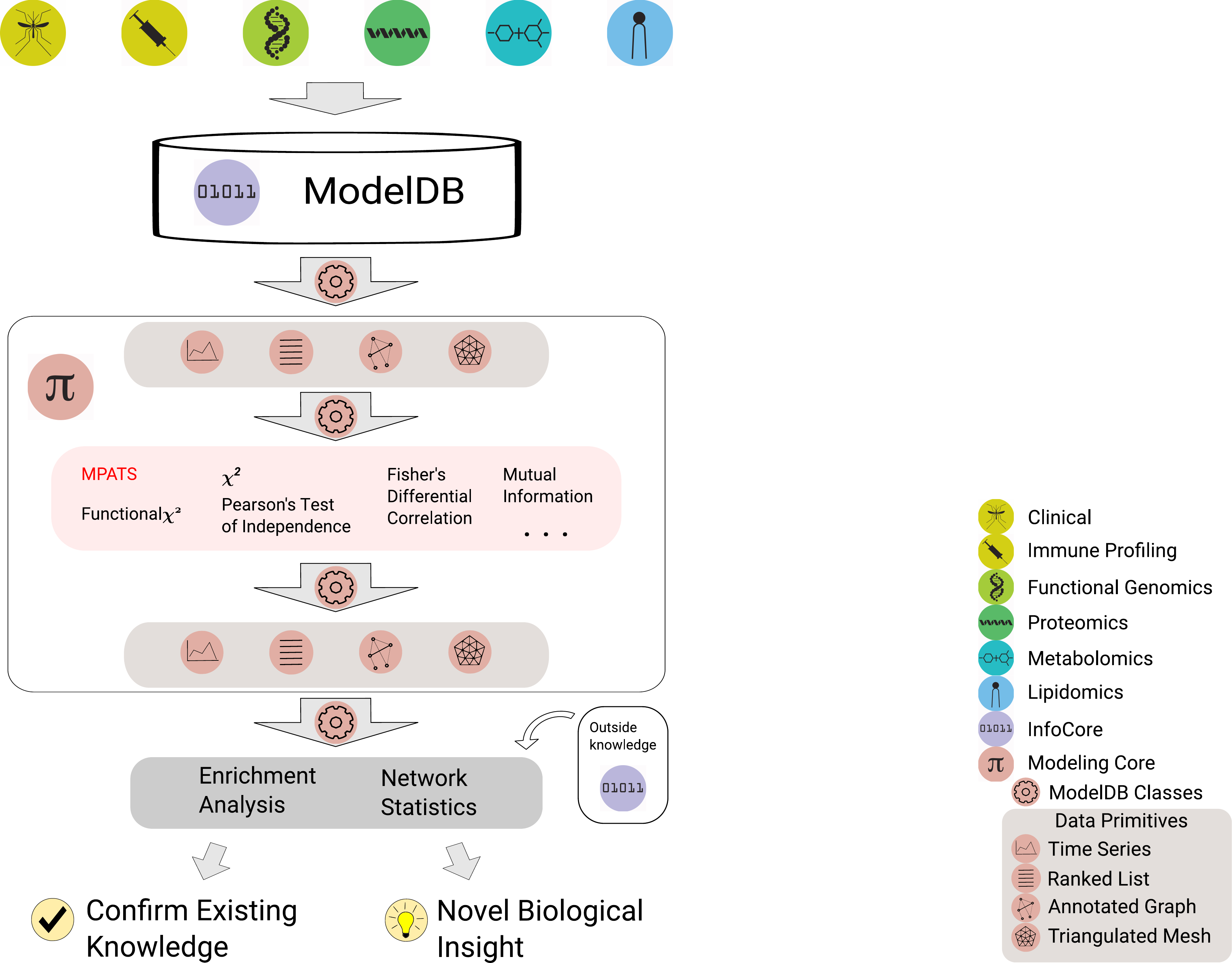}
 		\caption{Multi-omic data analysis pipeline }
 		\label{fig:flowdiagram}
 	\end{figure}

A multi-omic systems biology study of \textit{Plasmodium cynomolgi} infection in five \textit{Macaca mulatta} was undertaken as described by Joyner et al \cite{joyner2016plasmodium}.  Metabolomic, clinical, and cellular measurements were made daily, while transcriptomic, cytokine, and immune profiling measurements were taken at only seven time points over the course of the experiment. In this experiment, two subjects developed severe malaria while two subjects only experienced mild malarial symptoms.  The computational challenge was to integrate this data from multiple time scales and multiple -omic measurements into a biologically relevant conclusion about the molecular and cellular mechanisms differentiating mild vs severe malaria. Figure \ref{fig:flowdiagram} summarizes the data analysis and integration steps.

As described in Figure \ref{fig:flowdiagram}, data from multiple types of measurements were first converted into data primitives and stored in a relational database (called ModelDB here) for efficient retrieval.  Once the data had been accessed from the database, data primitives were used for the intermediate steps in all calculations, enabling the most significantly changed variables across the data types to be identified.  The results of this analysis are described by Yan et al (submitted).  

\section{Conclusions}\label{sec:Conclusion}

Rather than focus on the integration of only one or two data types, data primitives allow the integration of multiple data types in a modular, extensible fashion.  Data primitives are the foundation of the SKED framework, in which data primitives are used for data storage to allow integration of large, heterogeneous data sets and increase reproducibility and reliability of computational analyses.
		
\section{Abbreviations}\label{sec:Abbreviations}
\textbf{SKED: }Scientific Knowledge Extraction from Data

\section{Declarations}\label{sec:Declarations}

\textbf{Acknowledgements}

\textbf{Funding}

This project was funded in part by Federal funds from the US National Institute of Allergy and Infectious Diseases, National Institutes of Health, Department of Health and Human Services under contract \#HHSN272201200031C, which supports the Malaria Host-Pathogen Interaction Center (MaHPIC).

\textbf{Availability of data and materials}

A public interface featuring the MaHPIC dataset will soon be available at \texttt{http://euler.math.uga.edu}.

\textbf{Authors’ contributions}

All authors have read and approved the final version of the manuscript.



\textbf{Competing interests}

The authors declare that they have no competing interests.

\textbf{Consent for publication}

Not applicable.

\textbf{Ethics approval and consent to participate}

Not applicable.


\end{document}